\documentclass[aps,twocolumn,showpacs,nofootinbib]{revtex4}
\usepackage{graphicx}

\newcommand{\beq}{\begin{equation}}
\newcommand{\eeq}{\end{equation}}
\newcommand{\beqa}{\begin{eqnarray}}
\newcommand{\eeqa}{\end{eqnarray}}

\def\<{\langle}
\def\>{\rangle}
\def\ket#1{|#1\rangle}
\def\bra#1{\langle\, #1\,|}

\newcommand{\complex}{{\kern .1em {\raise .47ex\hbox {$\scriptscriptstyle |$}}\kern -.4em {\rm C}}}
\newcommand{\real}{{{\rm I} \kern -.19em {\rm R}}}
\def\opone{\leavevmode\hbox{\small1\normalsize\kern-.33em1}}

\begin{document}

\title{Bell inequalities: many questions, a few answers}

\author{Nicolas Gisin}

\affiliation{
    Group of Applied Physics, University of Geneva, 1211 Geneva 4,
    Switzerland}

\date{\today}

\begin{abstract}
What can be more fascinating than {\it experimental metaphysics}, to quote one of Abner Shimony's
enlightening expressions? Bell inequalities are at the heart of the study of nonlocality. I present
a list of open questions, organised in three categories: fundamental; linked to experiments; and
exploring nonlocality as a resource. New families of inequalities for binary outcomes are
presented.
\end{abstract}

\maketitle

\section{Introduction}\label{intro}
This Festschrift in honor of Abner Shimony is the ideal occasion to review some of the many
questions about Bell inequalities that remain open, despite more than four decades of active
research and a vast number of publications on this fascinating subject. Indeed, Abner was - in
modern terminology - an early adaptor of the product {\it Bell inequality}. At that time, in the
1960's and 1970's, it required quite some courage and independence of thought, two qualities
characterizing Abner, to recognize the value of Bell's work on the foundations of Quantum Physics.
Even in the 1980's, after Aspect's experiments, Bell inequality was still considered a dirty work.
``Bohr sorted out all that years ago", was the standard answer. In those days, if you wanted your
work published in PRL or similar high-standard journals you had better avoid terms like Bell
inequality and (even worse) quantum nonlocality.

Starting with Artur Ekert's PRL relating Bell inequalities with quantum key distribution things
have drastically changed \cite{Ekert91}. Today it would be hard to find an issue of PRL without a
mention of Bell inequality, nonlocality and - on top of it all - ``the potential relevance of the
presented work for quantum information processing". It is nice to see how human physicists are! And
who is more human, in the most noble sense of the word, than Abner? Abner, you helped me
tremendously; moreover, you did so at a time when I really needed it. Thank you Abner!

Let's return to the product {\it Bell inequalities}. Today it is fashionable, see Fig. 1, although
I suspect that a large majority of physicists would still be unable to properly derive any Bell
inequality. I bet that in a few decades Bell inequalities will be taught at high school, because of
their mathematical simplicity, their force as an example of the scientific methodology and their
huge impact on our world view. Yet, there remains a surprisingly large number of open questions,
several of which are listed in section \ref{questions}. Section \ref{ASBell} presents a new family
of Bell inequalities for an arbitrary even number of settings and binary outcomes. In appendix B an
elegant Bell inequality for qubits is presented; its optimal quantum violation requires
measurements of all three Pauli matrices $\sigma_x$, $\sigma_y$ and $\sigma_z$. However, let's
start in section \ref{BellIneq} by defining the notation.

\begin{figure}
\includegraphics[width=100mm,scale=0.35]{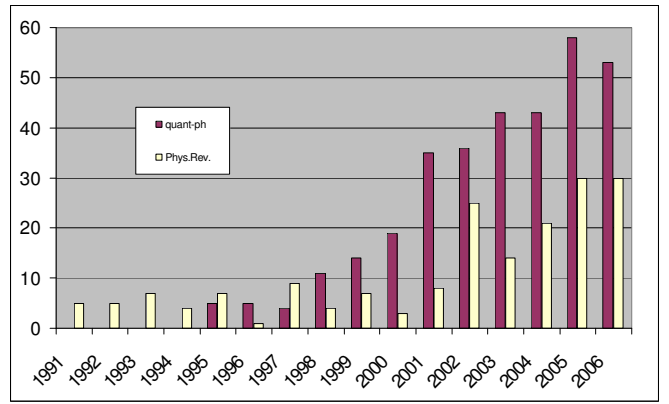} \caption{(Color online) Number of
occurrences of the words {\it Bell inequality} or {\it Bell inequalities} in the title or abstract
of papers published during the last 16 years on the quant-ph preprint server and in Physical Review
(PRL+PRA+PRB+PRC+PRD+PRE).}\label{fig1}
\end{figure}

\section{Bell inequalities}\label{BellIneq}
Bell inequalities are relations between conditional probabilities valid under the locality
assumption. Hence, {\it a priori} they have nothing to do with quantum physics (and thus should not
be written using quantum operators). However, it is the fact that quantum physics predicts a
violation of these relations that makes them interesting. The purpose here is not to present yet
another derivation of Bell inequalities, but merely to fix notation. Let $p(a,b,c,...|x,y,z,...)$
denote the conditional probability that players $A,B,C,...$ produce the outcome $a,b,c,...$ when
they receive the input $x,y,z,...$. Typically the players are physicists that perform measurements
$x,y,z,...$ with results $a,b,c,...$. Note that $a,b,c,...$ need not be numbers. We call the
conditional probabilities $p(a,b,c,...|x,y,z,...)$ correlations. We assume the numbers of players,
inputs and outcomes are all finite. Under the assumption of locality [i.e. there is a probability
distribution $p(\lambda)$ such that $p(a,b,c,...|x,y,z,...)=\sum_\lambda p(\lambda)\cdot
p(a|x,\lambda)\cdot p(b|y,\lambda)\cdot p(c|z,\lambda)\cdot...$] the set of all correlations is
convex with finitely many vertices. Such sets are called polytopes \cite{Pitowski89}. Thus, for any
given finite number of players, inputs and outcomes, the set of local correlation
$p(a,b,c,...|x,y,z,...)$ is called the {\it local polytope} \cite{Pitowski89}. These polytopes are
bounded by facets (hyperplanes). Each facet can be described by a linear equation:
$\sum_{a,b,c,...,x,y,z,...} C_{abc...}^{xyz...} p(a,b,c,...|x,y,z,...) = S_{lhv}$ with real
coefficients $C_{abc...}^{xyz...}$ and $S_{lhv}$. All local correlations lie on one side of the
facet, hence they necessarily satisfy the inequality:
\beq
\sum_{a,b,c,...,x,y,z,...} C_{abc...}^{xyz...} p(a,b,c,...|x,y,z,...) \leq S_{lhv}
\eeq
Such inequalities are called tight Bell inequalities (for an elegant, but not tight Bell
inequality, see appendix B). We say that a quantum state $\rho$ is nonlocal iff there are
measurements on $\rho$ that produce a correlation that violates a Bell inequality.

The famous CHSH inequality \cite{CHSH} reads
\beqa
E(x=0,y=0)&+&E(x=0,y=1)+ \nonumber\\
E(x=1,y=0)&-&E(x=1,y=1)\leq2
\eeqa
where in our notations $E(x,y)=p(a=b|x,y)-p(a\neq b|x,y)$. It is convenient to use the following
self-explanatory matrix notation:
\beq
CHSH\doteq\left(\matrix{
  +1 & +1 \cr
  +1 & -1 \cr
}\right)\le 2
\eeq
This CHSH inequality is the only tight Bell inequality for the bipartite case (i.e. two players)
with binary inputs and outcomes (up to local symmetries).

Let us emphasize that the entire game consists for each player in producing, for of a given
situation, a classical outcome with some probability for any possible input. In the quantum case
this implies performing measurements with classical outcomes on a given quantum state $\rho$.
Accordingly, the players can't combine several instances, i.e. several quantum states $\rho$, and
perform quantum information processing on them, i.e. exploit coherent measurement on $\rho^{\otimes
n}$ for $n\ge2$. Note that this does not exclude the situation where the players receive a fixed
number of states, like e.g. $\rho^{\otimes 3}$, but this is a different game from the one based on
$\rho$. Clearly, {\it a priori} a state $\rho$ can be local, while $\rho^{\otimes n}$ is nonlocal
for all $n\ge n_{threshold}>1$.

\section{Open questions}\label{questions}
The open questions can be organized in three groups. First, the fundamental questions, most in the
spirit of Bell. Next, questions more related to experiments, in the spirit of Abner's works (e.g.
the famous CHSH-Bell inequality and the detection loophole). Finally, Bell-like inequalities for
nonlocal resources, the most timely research on nonlocality.

Note that many open questions in quantum information theory are listen on the web page
\cite{QuestionsWerner}.

\subsection{fundamental questions}
There are infinitely many Bell inequalities. Even if one is restricted to tight Bell inequalities
corresponding to facets of the polytope of local correlations, the number of Bell inequality is
infinite. Restricting the given number of inputs and outcomes limits the number of Bell
inequalities, but it is a computationally hard problem to list them \cite{Pitowski89}.

\begin{enumerate}
\item Why is the CHSH inequality almost always the most efficient one to prove a quantum state to be nonlocal?
Until 2004 there was no example of a quantum state not violating the CHSH inequality, but violating
some other Bell inequality \cite{Collins04}. Still today, no natural example, i.e. a state with
some natural symmetry, has been found. This leads to the concept of {\it relevant Bell
inequalities}: an inequality is relevant with respect to a given set of inequalities if there is a
quantum state violating it, but not violating any of the inequalities in the set.

\item Is there a finite set of inequalities such that no other inequality is relevant with respect
to that set? What if one limits the dimension of the Hilbert space?

\item Find an inequality that is more efficient than the CHSH one for the Werner states \cite{Werner89} or prove it
is impossible. In dimension two, Werner states are simply mixtures of a maximally entangled pure
state $\psi$ with noise (i.e. the identity operator): $\rho_W=W\ket{\psi}\bra{\psi}+(1-W)\opone/4$,
where $W$ is the visibility. A local model exists for $W\lesssim 0.66$ \cite{Groetendick06}, the
CHSH inequality proves Werner states to be nonlocal for $W>1/\sqrt{2}$. The region in between is
unknown. The same question for the isotropic state (mixture of maximally pure state and noise) has
been answered in part in \cite{Kaszlikowskietal00,CGLMP02} where a generalization of the CHSH
inequality to arbitrary numbers of outcomes has been shown to be more efficient. But for the
isotropic states there remains also a gap in between the best known local model
\cite{AcinIsotropic} and the proven nonlocality visibility threshold.

\item Is hidden nonlocality generic for all entangled quantum states, including mixed states?
In dimension $\geq5$, Popescu proved that the Werner states, although admitting local models, have
hidden nonlocality, i.e. there are local filters such that if the Werner state passes the filters,
then the resulting state violates the CHSH inequality \cite{Popescu95}, see also \cite{Gisin96} for
a simple example of hidden nonlocality. In the same vein, one should ask whether for all quantum
states with hidden nonlocality there is a Bell inequality, possibly with more inputs and outcomes,
that can be violated by this state? Finally, is there an example of hidden nonlocality that
requires a sequence of local filters rather than a single one (the local model should reproduce all
intermediate results)?

\item Prove some entangled quantum states to be local. This requires one to prove the existence of a
local model. This has been done for Werner states (see \cite{Werner89} for projective measurements
and \cite{Barrett02} for general POVMs) and very recently for isotropic states
\cite{AcinIsotropic}. A weaker form of this question asks for a proof that a state can't violate
any Bell inequality with less than a given number of inputs and/or outcomes. There is only one
general result to this question, see the elegant construction in \cite{Therhal03}.

\item Why are almost all known Bell inequalities for more than 2 outcomes maximally violated by states that are not
maximally entangled \cite{Acin02}? There is quite a lot of evidence that entanglement and
nonlocality are different resources \cite{MethotScarani06}.

\item Can all Bell inequalities with $d$ outcomes be maximally violated by a quantum state of
dimension $d$? Or is there an example requiring states of dimension larger than the number of
outcomes? In reference \cite{Helle03}, a Bell inequality with $m$ outcomes on Alice's side and
binary outcomes on Bob's side is presented. It is maximally violated by the maximally entangled
state in dimension $m$.

\item Is there a local quantum state $\rho$ such that $\rho^n$ violates some Bell inequality? Note
that if the state $\rho$ is distillable, then $\rho^n$, for large enough $n$, contains hidden
nonlocality.

\item Find genuine n-party inequalities violated by all n-party pure entangled states.
In the case of two parties, the CHSH inequality is such an example, i.e. it can be violated by any
pure entangled state of whatever dimension \cite{Gisin91,GisinPeres92}. In the case of three
parties there are entangled states that do not violate the MBK inequality
\cite{Mermin90,BelinskiiKlyskho93}. In \cite{Acinetal04, Chen04}, a Bell inequality is presented
that shows numerical evidence that all 3-party pure entangled state violate it. But the case of
arbitrarily many parties is still open. Note that all n-party pure entangled states can always be
projected onto a 2-party pure entangled state by projecting n-2 parties onto appropriate local pure
states \cite{PR92}. This can be formulated as a tight Bell inequality where n-2 parties have only a
single input. Hence, there is a set of
$\left(%
\begin{array}{c}
  n \\
  n-2 \\
\end{array}%
\right)$ inequalities that does the job. But is there a single inequality?

\item There is no known Bell inequality that requires POVMs for optimal violation on some quantum
states. For binary outcomes, one can prove that POVMs are never relevant \cite{CleveTonner04}, but
for larger a number of outcomes the question is open.

\item Almost all Bell inequalities are maximally violated by quantum states and measurements that can all be written,
in an appropriate basis, using only real numbers. This is surprising since interference, a basic
quantum property, ``requires" complex numbers. It would be nice to find Bell inequalities suitable
for distinguishing real Hilbert spaces from complex ones (i.e. an inequality that can only be
violated by states and settings that require complex numbers). An example is \cite{Helle03}.

\item Is there a bound entangled state that violates some Bell inequality? In \cite{Masanes06}
Masanes proves that no bound entangled state violates the CHSH inequality. But what about other
Bell inequalities? Note that in the case of 3 players or more, it is important to distinguish
different meanings of bound entanglement: bound means that the players can't distill a maximally
entangled states between all of them; while totally bound means that even if some parties join into
groups, they still can't distil entanglement between the groups. D\"urr found a bound entangled
state of 8 qubits that violates the MKB inequality \cite{Dur01}. However the violation is small,
indicating that there is no 8-party entanglement \cite{GisinBechmann98}. Actually it was then
demonstrated for qubits that any violation of a Bell inequality, with 2 inputs per player implies
that the players, can join into groups such that the groups can distill a maximally entangled state
\cite{Acin01,ASW02}.

\item In the case of more than 2 parties, find inequalities testing models that assume bi-partite
nonlocality but no arbitrary multi-partite nonlocality. A first example was presented already in
1987 by Svetlichny \cite{Svetlichny87} and generalized in
\cite{Collins_Svetlichny02,SeevinckSvetlichny02,Cereceda02}.

\item Find families of Bell inequalities valid for any number of inputs and outcomes. An example of such a family is
presented in \cite{Collins04}. Another example is presented in this paper, see section
\ref{ASBell}, though valid only for binary outcomes and even numbers of settings. The MKB
inequality \cite{Mermin90,BelinskiiKlyskho93} is an example of a family of Bell inequalities with
fixed numbers of inputs and outcomes, but for arbitrarily many parties. See also the recent
\cite{Nagata06}.

\item Given a multi-party quantum state $\rho$, how can one know whether $\rho$ is nonlocal, i.e. whether
there is a Bell inequality and measurements such that quantum physics predicts a violation of the
inequality? For pairs of qubits and the CHSH inequality this problem has been solved in 1995 by the
Horodecki family \cite{Horodecki95}, but the general problem seems exceedingly hard.

\end{enumerate}

\subsection{Questions relevant for experiments}
The original Bell inequality \cite{Bell64} is, strictly speaking, not a Bell inequality according
to the modern terminology that we use here. Indeed, the original inequality required, besides
locality, another assumption about perfect correlations. Abner immediately recognized that this
auxiliary assumption made the entire enterprize non testable and searched for an inequality
involving only measurable quantities. This led him and his co-workers to find the CHSH inequality.
Interestingly, the CHSH paper \cite{CHSH} already mentions the detection loophole, again underlying
the importance the authors gave to the experimental issues. Concerning the detection loophole, see
also \cite{Pearle70}.

\begin{enumerate}

\item Find Bell inequalities easier to test experimentally with today's technology, while avoiding all known loopholes.
Quantum nonlocality is so fundamental for our world view that it deserves to be tested in the most
convincing way. It is thus surprising and annoying that no experiment to date has managed to close
simultaneously the locality loophole (space-like separation from the choice of settings until the
the classical data are secured) and the detection loophole. The latter consists in assuming that
the detection efficiency is independent of the hypothetical local variables (for example, if
polarization would be unknown, one would assume that all detectors are polarization insensitive, a
clearly wrong assumption). Reference \cite{GisinGisin99} presents a simple model reproducing all
quantum correlations on maximally entangled qubits assuming detection efficiencies of 2/3 (and
projective measurements). Violation of the CHSH inequality requires detection efficiencies of at
least 82.84\%, for maximally entangled states. There is only a single known inequality with few
settings that does better, though only marginally better, $\eta_{threshold}=\sqrt{2/3}\approx
81.65\%$ \cite{Pironio02}. This inequality has 3 settings on each side and is {\it not} a facet of
the polytope of local correlations. For numbers of settings larger than one hundred a better
inequality has been derived from communication complexity arguments \cite{BuhrmannMassar03}.
Interestingly, Philippe Eberhard noticed that partially entangled states are less sensitive to the
detection loophole \cite{Eberhard93}.

\item A timely variation of the previous question addresses situations where the detection efficiency
differs from one side of the experiment to the other. This is natural for experiments on
entanglement between quantum systems of different kinds, like e.g. an atom and a photon
\cite{Weinfurter06,Cabello07,Brunneretal07}.

\item Find inequalities suitable for a Bell test with simple quantum-optics states and homodyne detectors.
Indeed, the homodyne detection technique is well developed and always produces an outcomes. But
simple cases likes, e.g. a delocalized photon in state $\ket{0,1}+\ket{1,0}$, although clearly
entangled, does not violate the CHSH inequality with homodyne detection and a simple binarisation
of the measurement results. More complicated states could violate the CHSH, but only by a tiny
amount, see \cite{CerfGrangier04} and references therein.

\item Find inequalities for many settings. Experimentally one rarely measures precisely the four
probabilities that appear in the CHSH inequality. Most of the time a series of points is measured
and fitted with a sinus. Hence, an inequality for such series of points could be more appropriate.
Examples are given in \cite{Zukowski93,Gisin99} and in section \ref{ASBell}.

\end{enumerate}

\subsection{Bell-like inequalities for nonlocal resources}
This subsection presents recently opened questions and moves away from the traditional work on Bell
inequalities. It starts by admitting quantum nonlocality and aims at better quantifying it and at
understanding it as a new kind of resource. These questions investigate nonlocal but non-signaling
correlations \cite{Barrett05}. Recall that a correlation $p(a,b,c,...|x,y,z,...)$ is non-signaling
iff all the marginals are independent of the other players's inputs:
$\sum_{b,c,...}p(a,b,c,...|x,y,z,...)=p(a|x)$, $\sum_{a,c,d,...}p(a,b,c,...|x,y,z,...)=p(b,|y)$,
etc.

Bell inequalities are tests for correlations that can be simulated using only local resources and
shared randomness (a modern terminology for the obsolete {\it local hidden variables}). This view
raises the question of correlations that can be simulated using, in addition to shared randomness,
some finite amount of some given nonlocal resource. For example, it is known that any pair of
projective (Von Neumann) measurements on any maximally entangled state of two 2-level quantum
systems can be simulated using only shared randomness and a single PR-box (a sort of unit of
nonlocality) \cite{PR94,CerfGisinMassar05,BarrettPironio05,Dupuisetal07}. Hence, it is interesting
to characterize all correlations that can't be simulated using shared randomness and one PR-box.
Surprisingly, some correlations resulting from quantum measurements on partially entangled 2-level
systems are of that kind.

\begin{enumerate}
\item Is there a Bell-like inequality valid for all correlations simulable with a single bit of
communication and violated by some partially entangled 2-qubit states? Actually, the entire field
of research considered in this subsection started with a paper presenting Bell-like inequalities
valid for 1 bit of communication \cite{BaconToner03}. However, the presented inequalities can't be
violated by any 2-qubit states. We know that maximally entangled 2-qubit states can be simulated
with a single bit of communication; thus such states don't violate any of the considered Bell-like
inequalities. However, the question remains open for partially entangled states.

\item Are all partially entangled qubit pairs not simulable by a single PR-box? A few Bell-like
inequalities satisfied by all correlations simulable by a single PR-box and shared randomness are
known \cite{Brunner05,Brunner06}. From these one knows that very poorly entangled states can't be
simulated with one PR-box, but the case of high-but-not-maximally entangled states is open.

\item Find inequalities satisfied by all correlation that can be simulated by two PR-boxes. Two
bits of communication suffice to simulate any two qubit state. Is the same true for two PR-boxes?

\item Find any non-signaling box \cite{Barrett05} with finitely many inputs and outcomes with which one can simulate
partially entangled states.

\item Find the {\it Quantum-Bell inequalities} that bound the correlations achievable with quantum
measurements and states? An example is the Tsirelson bound \cite{Tsirelson80} stating that quantum
correlations can't violate the CHSH inequality by more than the well known factor $2\sqrt{2}$, see
also \cite{Acin06}.

\item Can a secret key be distilled out of any nonlocal correlation, (secret against any
non-signaling adversary performing arbitrary individual attacks)
\cite{BarrettKentetal05,Scaranietal06,Acinetal06}? This question may appear to move away from {\it
Bell questions}, but it concerns the power of the nonlocal resources as witnessed by Bell
inequalities. It also addresses the question of the existence of bound information
\cite{boundInfo1,boundInfo2}, a classical analog to bound entanglement.

\end{enumerate}

\section{The AS-Bell inequality family}\label{ASBell}
I know of only a single family of bipartite Bell inequalities valid for any number of inputs and
outcomes \cite{Collins04}. In this section I briefly present a new family of bipartite Bell
inequalities for any even number of inputs and binary outcomes. I found this family by looking for
correlation Bell inequalities with a few inputs and binary outcomes. Recall that a correlation
inequality involves only expectation values: $E(x,y)=p(a=b|x,y)-p(a\neq b|x,y)$. For binary inputs,
the CHSH is the only inequality. For ternary inputs, there is no new correlation inequality
\cite{Collins04,Sliwa03}. For 4 inputs on each side, I searched numerically all possibilities
assuming small integer coefficient. I found only two new inequalities (the coefficient in the
matrix indicate the coefficients of the corresponding expectation values):
\beq
AS_4\doteq\left(\matrix{
  +1 & +1 & +1 & +1 \cr
  +1 & +1 & +1 & -1 \cr
  +1 & +1 & -2 & 0 \cr
  +1 & -1 & 0 & 0}\right)\le6
\eeq
\beq\label{D4}
D_4\doteq\left(\matrix{
  +2 & +1 & +1 & +2 \cr
  +1 & +1 & +2 & -2 \cr
  +1 & +2 & -2 & -1 \cr
  +2 & -2 & -1 & -1}\right)\le10
\eeq
Avis and co-workers demonstrated that these are indeed the only correlation inequalities for 4
inputs \cite{Ito06}. Inspired by inequality $AS_4$, it is not difficult to guess the form of the
next inequalities:
\beq
AS_6\doteq\left(\matrix{
  +1 & +1 & +1 & +1 & +1 & +1 \cr
  +1 & +1 & +1 & +1 & +1 & -1 \cr
  +1 & +1 & +1 & +1 & -2 & 0 \cr
  +1 & +1 & +1 & -3 & 0 & 0 \cr
  +1 & +1 & -2 & 0 & 0 & 0 \cr
  +1 & -1 & 0 & 0 & 0 & 0}\right)\le12
\eeq

\beq
AS_8\doteq\left(\matrix{
  +1 & +1 & +1 & +1 & +1 & +1 & +1 & +1 \cr
  +1 & +1 & +1 & +1 & +1 & +1 & +1 & -1 \cr
  +1 & +1 & +1 & +1 & +1 & +1 & -2 & 0 \cr
  +1 & +1 & +1 & +1 & +1 & -3 & 0 & 0 \cr
  +1 & +1 & +1 & +1 & -4 & 0 & 0 & 0 \cr
  +1 & +1 & +1 & -3 & 0 & 0 & 0 & 0 \cr
  +1 & +1 & -2 & 0 & 0 & 0 & 0 & 0 \cr
  +1 & -1 & 0 & 0 & 0 & 0& 0 & 0 }\right)\le20
\eeq

The generalization to arbitrary even number of inputs is straightforward. Note that $AS_2$ is
nothing but the CHSH inequality. Numerically, these $AS_n$ inequalities are tight and maximally
violated by maximally entangled qubit states for visibilities larger than $V_n$, with
$V_2=1/\sqrt{2}\approx0.7071$, $V_4\approx0.7348$, $V_{10}\approx0.7469$, $V_{32}\approx0.7497$,
$V_{50}\approx0.7499$. Apparently $V_\infty \approx 0.75$; this contrasts with the $I_{nn22}$
family presented in \cite{Collins04} where for binary outcomes and large numbers of inputs the
threshold visibility appears to tends to 1. All settings can be chosen to lie on a grand circle of
the Poincar\'e sphere.

\section{conclusion} \label{concl}
We are lucky to live at the time where physics discovers and explores the nonlocal characteristics
of Nature. Contrary to the nonlocality of Newtonian gravitation, quantum nonlocality is with us for
ever \cite{NewtonNonlocality,Gisin05}. Future historians of Science will describe our epoch as that
of the great discovery of nonlocality. The name of Abner Shimony will forever be associated with
this fascinating epoch.

The choice of questions listed in this contribution to Abner's Festschrift is necessarily somewhat
subjective. Others may like to add their favorite ones or to formulate the questions differently.
Important is the fact that there are many interesting open questions of very different kinds. The
basic maths is simple, but a deeper understanding requires concepts ranging from combinatorial and
complexity theories to algebra and geometry in high dimensions. Hence, it is likely that most of
the listed problems are hard. But their solutions, even partial solutions, will be valuable
contributions to one of the most fascinating research fields of the 21st century.


\section*{Acknowledgment}
This work has been supported by the EC under project QAP (contract n. IST-015848) and by the Swiss
NCCR {\it Quantum Photonics}. Thanks are due to Rob Thew, Toni Acin, Andr\'e M\'ethot, Sandu
Popescu and Valerio Sacarani for their comments on previous versions of this paper.

\section*{Appendix A: Some diagonal Bell inequalities}
Correlation Bell inequalities of a form similar to D4 (eq. \ref{D4}) can easily be found
numerically. For 5 inputs on each side there seems to exist only two such inequalities (at least I
found only two). They are entirely defined by their first line and the permutation rule as in
(\ref{D4}) (from one line to the next: shift each entry to the left, the entry that falls out is
re-introduced on the right hand side with the opposite sign):
\beqa
D5_1&\doteq&(1 1 0 1 1)\le8 \\
D5_2&\doteq&(3 2 1 1 3)\le20
\eeqa
For 6 inputs I found:
\beqa
D6_1&\doteq&(1 0 1 0 1 1)\le10 \\
D6_2&\doteq&(3 1 1 1 2 4)\le28 \\
D6_3&\doteq&(4 2 2 1 2 5)\le36 \\
D6_4&\doteq&(4 2 2 1 3 6)\le42
\eeqa
For more inputs, the numbers of such D-inequalities seems to grow rapidly.

\section*{Appendix B: An elegant Bell inequalities}
In ref. \cite{Helle03} Helle Bechmann-Pasquinucci and myself presented a Bell inequality tailored
for quantum cryptography in high dimension Hilbert spaces. Since this inequality seems to have a
few original features, like being optimally violated by states and quantum measurements requiring
complex numbers and Hilbert spaces of dimension larger than the number of outcomes on Bob's side
(but equal to the number of outcomes on Alice's side), I recall it in this appendix with the
notations used throughout this contribution. Moreover, this new way of looking at this inequality
underlines its similarity with communication complexity \cite{Brassard01}.

In this game, Alice receives as input a number $x\in \{0,1,...,n-1\}$, while Bob's input consists
of $n$ numbers $y_0,y_1,...,y_{n-1}$ with each $y_j\in \{0,1,...,m-1\}$. Basically, the goal is
that Alice outputs $a=y_x$. As such this would be merely an example of a communication complexity
game. But in our game, Bob can use a joker and refuse that this instance of the game counts.
Accordingly, Bob's outcome is binary. Whenever $b=0$, the score is null, whatever Alice's outcome.
Whenever $b=1$ the score is +1 if $a=y_x$ and -1 if $a\ne y_x$. Explicitly, the Bell inequality
reads:
\beqa
SHB=  \nonumber
\eeqa
\beqa
\sum_{\begin{array}{c}
  x=0...n-1 \\
  y_0...y_{n-1}=0...m-1 \\
\end{array}}
\begin{array}{c}
  \big( p(a=y_x, b=1|x,y_0,...,y_{n-1}) \\
  -p(a\ne y_x, b=1|x,y_0,...,y_{n-1})\big) \\
\end{array} \nonumber
\eeqa
\beq
\le S_{local}
\eeq
The optimal local strategy consists of Alice and Bob agreeing in advance on a sequence
$y_0^g,...,y_{n-1}^g$ and Alice producing $a=y_x^g$ while Bob accepts the game only for the inputs
$y_0,...,y_{n-1}$ for which the averaged score is positive:
\beq
S_{local}=\sum_{r=0}^{[\frac{n-1}{2}]} (n-2r) \left(%
\begin{array}{c}
  n \\
  r \\
\end{array}%
\right)
\eeq

Let us concentrate on the case $n=2$ for which $S_{local}=2$. The optimal quantum strategy requires Alice and Bob to share a maximally entangled state of dimension $m$. Alice measures her quantum
system in one out of two mutually conjugated bases, depending on her input $x=0$ or $x=1$. Bob
receives two symbols as input, $y_0$ and $y_1$, corresponding to two quantum states, one in each of
the two bases. He applies to his quantum system a measurement described by the projector onto the
state which lies precisely in between the two states that correspond to $y_0$ and $y_1$ (since the
two states belong to two mutually conjugated bases, such an intermediate state is always uniquely
define. Take for instance the eigenstate with maximal eigenvalue of the density matrix obtained by
a 50-50\% mixture of the two states). If Bob's projection is successful, this projects Alice's
state onto the state that maximizes her chance of finding the correct outcome. In such a case Bob
outputs $b=1$. In the alternative case, i.e. failure of his projection measurement, he outputs
$b=0$; that is, Bob's outcome is his measurement result. With this quantum strategy, Alice and Bob
beat the optimal local strategy by a factor $\sqrt{m}$:
\beq
S_{quantum}=2\sqrt{m}>S_{local}=2
\eeq
Note that for $m=2$, this reduces to the well studied CHSH inequality. Indeed, although in this
case Bob has formally four possible inputs, the corresponding four projectors form two bases.
Explicitly, Alice measures one of the two operators $\sigma_x$ or $\sigma_z$, depending on her
input, and Bob measures in the intermediate bases $\sigma_{+45^0}$ (for his inputs $0,0$ and $1,1$)
or $\sigma_{-45^0}$ (for inputs $0,1$ and $1,0$).

For $n=2$ and $m=3$ the quantum optimum of $2\sqrt{3}\approx 3.464$ is reached by the strategy
summaried above and presented in \cite{Helle03}. Numerical evidence suggests that if one restricts
oneself to settings that can be expressed using only real numbers, the maximum is slightly lower:
$10/3\approx 3.333$ \cite{Helle03}. Moreover, this maximum is reached for a non-maximally entangled
state. But it is unknown whether a higher score can be achieved using only real numbers in larger
Hilbert spaces.

The case $n=3$, $m=2$ appears also to be interesting. Indeed, the quantum maximum is
$4\sqrt{3}\approx 6.928$, while the maximum using only real numbers is reached by the singlet state
at $2+2\sqrt{5}\approx 6.472$. This might open the possibility to test correlations requiring complex Hilbert
spaces (however, here again it remains to test the inequality in higher dimensions).

Note that this inequality $n=3,m=2$ can also be written as a correlation inequality. Indeed, Bob's
$m^n=8$ inputs can be grouped into 4 projective measurements. In this form, this inequality reads:
\beq
S_{3\times 4}\doteq\left(\matrix{
  +1 & +1 & +1  \cr
  +1 & -1 & -1  \cr
  -1 & +1 & -1  \cr
  -1 & -1 & +1 }\right)\le6   \label{S3x4}
\eeq
\begin{figure}
\includegraphics[width=100mm,scale=0.35]{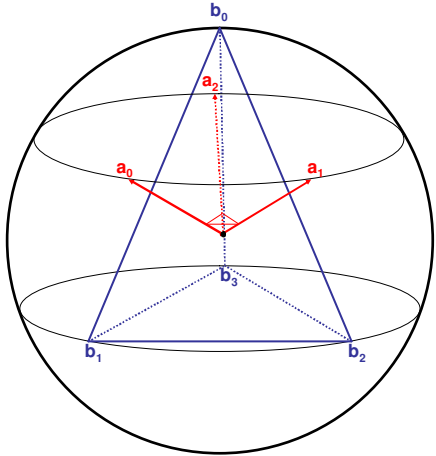} \caption{(Color online) Measurement settings represented on the
Poincar\'e sphere for the elegant inequality $S_{3x4}$ defined by eq. (\ref{S3x4}). Alice's 3
settings are represented by 3 mutually orthogonal vectors, and Bob's 4 settings by the vertices of
the tetrahedron.} \label{fig2}
\end{figure}

Another elegant feature of this case $n=3$, $m=2$ is seen when the optimal settings are represented
on the Poincar\'e sphere: for Alice the three vectors are mutually orthogonal, while Bob's four
vectors are on the vertices of the tetrahedron, see Fig. 2.

To conclude, let us note that most inequalities presented in this appendix, in particular the
elegant $S_{3\times 4}$, are not facets of the local polytope. This indicates that the geometry of
the local polytope doesn't match the symmetries of elegant quantum states and measurements. In the
case of 3 and 4 inputs on Alice and Bob's side, respectively, all facets are known
\cite{Collins04}, hence one shouldn't be surprised that the new inequality $S_{3\times 4}$ is not a
facet.


\end{document}